\documentclass{comnet}

\usepackage{amsmath} 
\usepackage{url} 
\usepackage{algorithm}
\usepackage{algorithmic}
\usepackage{graphicx}
\usepackage{hyperref}
\hypersetup{colorlinks=true, linkcolor=blue, filecolor=magenta, urlcolor=blue, citecolor=blue}
\AtBeginDocument{
  \label{CorrectFirstPageLabel}
  \def\fpage{\pageref{CorrectFirstPageLabel}}
}

\begin{document}

\title{fastball: A fast algorithm to randomly sample bipartite graphs with fixed degree sequences}

\shorttitle{fastball: algorithm to sample bipartite graphs} 
\shortauthorlist{K. Godard and Z. P. Neal} 

\author{
\name{Karl Godard}
\address{Electrical Engineering and Computer Science, University of Michigan}
\and
\name{Zachary P. Neal$^*$}
\address{Psychology Department, Michigan State University\email{$^*$Corresponding author: \href{mailto:zpneal@msu.edu}{zpneal@msu.edu}}}
}

\maketitle

\begin{abstract}
{Many applications require randomly sampling bipartite graphs with fixed degrees, or randomly sampling incidence matrices with fixed row and column sums. Although several sampling algorithms exist, the ``curveball'' algorithm is the most efficient with an asymptotic time complexity of $O(n~log~n)$, and has been proven to sample uniformly at random. In this paper, we introduce the ``fastball'' algorithm, which adopts a similar approach but has an asymptotic time complexity of $O(n)$. We show that a C\texttt{++} implementation of fastball randomly samples large bipartite graphs with fixed degrees faster than curveball, and illustrate the value of this faster algorithm in the context of the fixed degree sequence model for backbone extraction.}
{Bipartite, Configuration Model, Fixed Degree, Markov Chain, Randomization, Sampling}
\end{abstract}

\section{Introduction}
Many applications require randomly sampling bipartite graphs with fixed degrees, or randomly sampling incidence matrices with fixed row and column sums. In network science the sample provides an empirical null model for evaluating graph properties such as nestedness \cite{bruno2020ambiguity} and co-occurrence \cite{gotelli2000null,neal2014backbone,cimini2019statistical}. It also arises in other fields: in physics where the space of fixed-degree graphs can be viewed as a microcanonical ensemble representing a thermodynamic system \cite{barre2007ensemble,touchette2015equivalence,squartini2015breaking}; in mathematics where the sample can give insight into the cardinality of the space \cite{barvinok2010number}; and in quantitative psychology where it is useful for estimating Rasch models \cite{verhelst2008efficient}. These applications typically require drawing a large number of samples, therefore an efficient and unbiased sampling algorithm is essential.

We begin by formally stating the problem: Let $\mathcal{G}$ be the space of all bipartite graphs $\mathbf{G}$ containing $n$ top nodes with degrees $N = N_1, N_2 \dots N_n$ and $m$ bottom nodes with degrees $M = M_1, M_2 \dots M_m$. How can we randomly sample $\mathbf{G}~\in~\mathcal{G}$ with uniform probability? This question can also be formulated in matrix terms: Let $\mathbf{M}$ be the incidence matrix of $\mathbf{G}$, and $\mathcal{M}$ be the space of all $n \times m$ binary matrices with given row sums $N = N_1, N_2 \dots N_n$ and column sums $M = M_1, M_2 \dots M_m$. How can we randomly sample $\mathbf{M}~\in~\mathcal{M}$ with uniform probability? 

In this paper, we propose and demonstrate the ``fastball'' algorithm, which provides an efficient and unbiased method to sample $\mathbf{G}~\in~\mathcal{G}$ or $\mathbf{M}~\in~\mathcal{M}$. Fastball's asymptotic time complexity is $O(n)$ time, making it \textit{algorithmically more efficient} than the existing ``curveball'' algorithm, which has an asymptotic time complexity of $O(n~log~n)$. We use a numerical experiment to demonstrate that when implemented in a low-level language such as C\texttt{++}, fastball is also \textit{practically faster} than curveball for drawing samples of bipartite graphs. We use the example of extracting the backbone of a legislative co-sponsorship network to illustrate fastball's practical application in network science, where fastball draws the samples necessary to extract a signed network in 11 minutes compared to curveball's 27 minutes.

The remainder of the paper is organized in four sections. In section \ref{sec:background} we briefly review bipartite randomization algorithms, focusing on the curveball algorithm, which is currently the fastest. In section \ref{sec:fastball} we introduce the fastball algorithm as a more efficient randomization and sampling algorithm. In section \ref{sec:results} we compare the running time of fastball and curveball, then illustrate a practical application of fastball for extracting the backbone of bipartite projections. Finally, in section \ref{sec:discussion} we conclude by identifying directions for future research.

\section{Background}
\label{sec:background}
Several methods have been proposed for randomizing and sampling bipartite graphs and incidence matrices \cite{penschuck2020recent}. \textit{Fill} methods proceed by filling an initially empty graph by adding vertices via the configuration model \cite{blanchet2013characterizing}, or filling an initially empty matrix with 0s and 1s following the Gale-Ryser theorem \cite{gale1957theorem,ryser1957combinatorial}. Alternatively, \textit{swap} methods proceed by swapping (i.e. re-wiring) edges in a graph \cite{boroojeni2017generating}, or swapping checkerboard patterns (e.g., swapping $^0_1~^1_0$ with $^1_0~^0_1$) in a matrix. Other more sophisticated methods have been proposed that rely on sequential importance sampling \cite{admiraal2008networksis,chen2006sequential,verhelst2008efficient} or simulated annealing \cite{bezakova2007sampling}. 

The curveball algorithm is the current state-of-the-art, and is distinguished from these other methods in two ways. First, while it relies on the swap method, it performs multiple swaps simultaneously in each iteration called a `trade,' which makes it more efficient \cite{strona2014fast,carstens2018speeding}. Second, it has been proven to sample uniformly at random \cite{carstens2015proof}, which makes it unbiased.

\begin{figure}
\caption{Example of the curveball algorithm}
\label{fig:curveball}
\centering
\includegraphics[width=\textwidth]{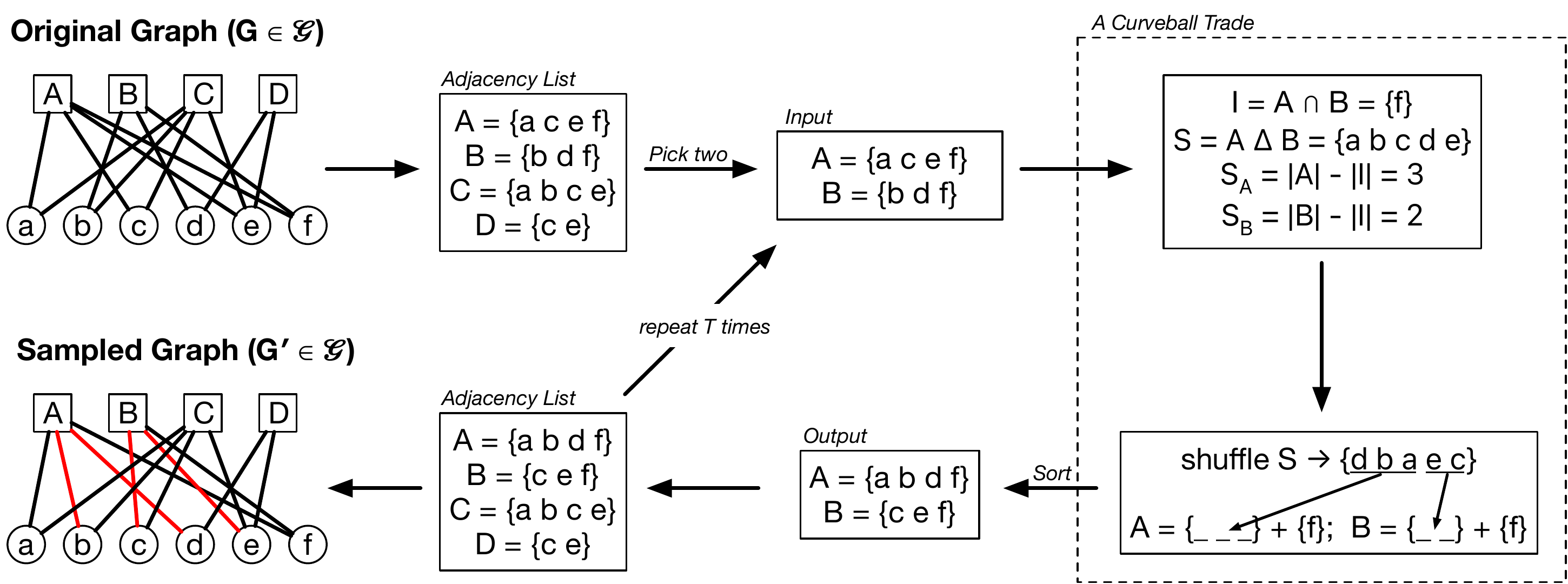}
\end{figure}

Figure \ref{fig:curveball} outlines the steps of the curveball algorithm, which illustrates how a starting bipartite graph $\mathbf{G}$ is randomized to yield a new graph $\mathbf{G}'$ that is randomly chosen from $\mathcal{G}$. The algorithm developers noted that this randomization process resembles how children may trade baseball cards, giving the algorithm its name, so we use this metaphor to make the example concrete \cite{strona2014fast}. The input is a bipartite graph $\mathbf{G}$. We use uppercase letters A--D to denote the top nodes (i.e., children), and lowercase letters a--f to denote the bottom nodes (i.e., baseball cards). For the purposes of the algorithm, $\mathbf{G}$ is represented as an \textit{adjacency list} that records each top node's neighbors (e.g., child A has cards a, c, e, and f). 

First, two top nodes $i$ and $j$ are randomly selected (e.g., children A and B will engage in card trading). Second, we find $i$ and $j$'s intersection $I$ (e.g., both children have card f), symmetric difference $S$ (e.g., only one child has cards a, b, c, d, and e), and cardinality of their contribution to the symmetric difference $S_i$ and $S_j$ (e.g., child A has 3 unique cards and child B has 2 unique cards). Third, we randomly shuffle $S$, and assign the first $S_i$ elements of $S$ plus $I$ to $i$ (e.g., child A now has cards a, b, d, and f), and the last $S_j$ elements of $S$ plus $I$ to $j$. Finally, top nodes $i$'s and $j$'s new adjacency lists are sorted and updated in the complete adjacency list, at which point the process can repeat. Using a card trading metaphor, this process mirrors two children placing all their unique cards in a pile, shuffling it, then randomly drawing the same number of cards from the pile that they put in.

The curveball algorithm offers two notable advantages over alternative approaches to randomizing $\mathbf{G}$ and sampling from $\mathcal{G}$. First, it has been proven to sample $\mathbf{G}~\in~\mathcal{G}$ uniformly at random \cite{carstens2015proof}. Second, it has been shown to mix more rapidly than swap methods because while swap methods change the position of only one edge in each iteration, a single curveball \textit{trade} can perform many such \textit{swaps} \cite{carstens2018unifying,carstens2018speeding}.

\begin{algorithm}[H]
\caption{Curveball Trade algorithm, $O(n~log~n)$}
\label{alg:curveball}
\begin{algorithmic}
\STATE \textbf{Input:} Sorted vector of $i$'s neighbors $N_i$ \& Sorted vector of $j$'s neighbors $N_j$\\
\STATE \textbf{Output:} Sorted vector of $i$'s neighbors $N'_i$ \& Sorted vector of $j$'s neighbors $N'_j$\\
\STATE 
\STATE Let $I = N_i \cap N_j \text{ and let } S = N_i~\Delta~N_j$ \hfill $O(n)$\\
\STATE Shuffle $S$ \hfill $O(n)$\\
\STATE Let $N'_i = I~+$ the first $|N_i| - |I|$ elements of $S$ \hfill $O(n)$\\
\STATE Let $N'_j = I~+$ the last $|N_j| - |I|$ elements of $S$ \hfill $O(n)$\\
\STATE Sort $N'_i$ \hfill $O(n~log~n)$\\
\STATE Sort $N'_j$ \hfill $O(n~log~n)$
\end{algorithmic}
\end{algorithm}

Recent work has improved curveball's efficiency by performing trades for multiple pairs of top nodes in parallel \cite{carstens2018unifying} and by using I/O-efficient techniques to manage the handling of adjacency lists \cite{carstens_et_al}. However, these innovations still rely on the same algorithm for performing each curveball trade, which we aim to improve. The algorithm for performing a curveball trade is shown in Algorithm \ref{alg:curveball}, which also shows the time complexity of each step. The input vectors of $i$'s and $j$'s neighbors must be sorted so that their intersection $I$ and symmetric difference $S$ can be found efficiently and simultaneously in $O(n)$ time. Because in practice curveball trades are performed repeatedly by the curveball algorithm, the output vectors must also be sorted so they are ready for the next trade, which requires $O(n~log~n)$ time.

\section{The fastball algorithm}
\label{sec:fastball}
The fastball algorithm is very similar to the curveball algorithm, but performs trades differently. In this section we first review a concrete example of the fastball algorithm, again using the baseball card trading metaphor, then present the more efficient fastball trade algorithm.

\begin{figure*}
\caption{Diagram and example of the fastball algorithm}
\label{fig:fastball}
\centering
\includegraphics[width=\textwidth]{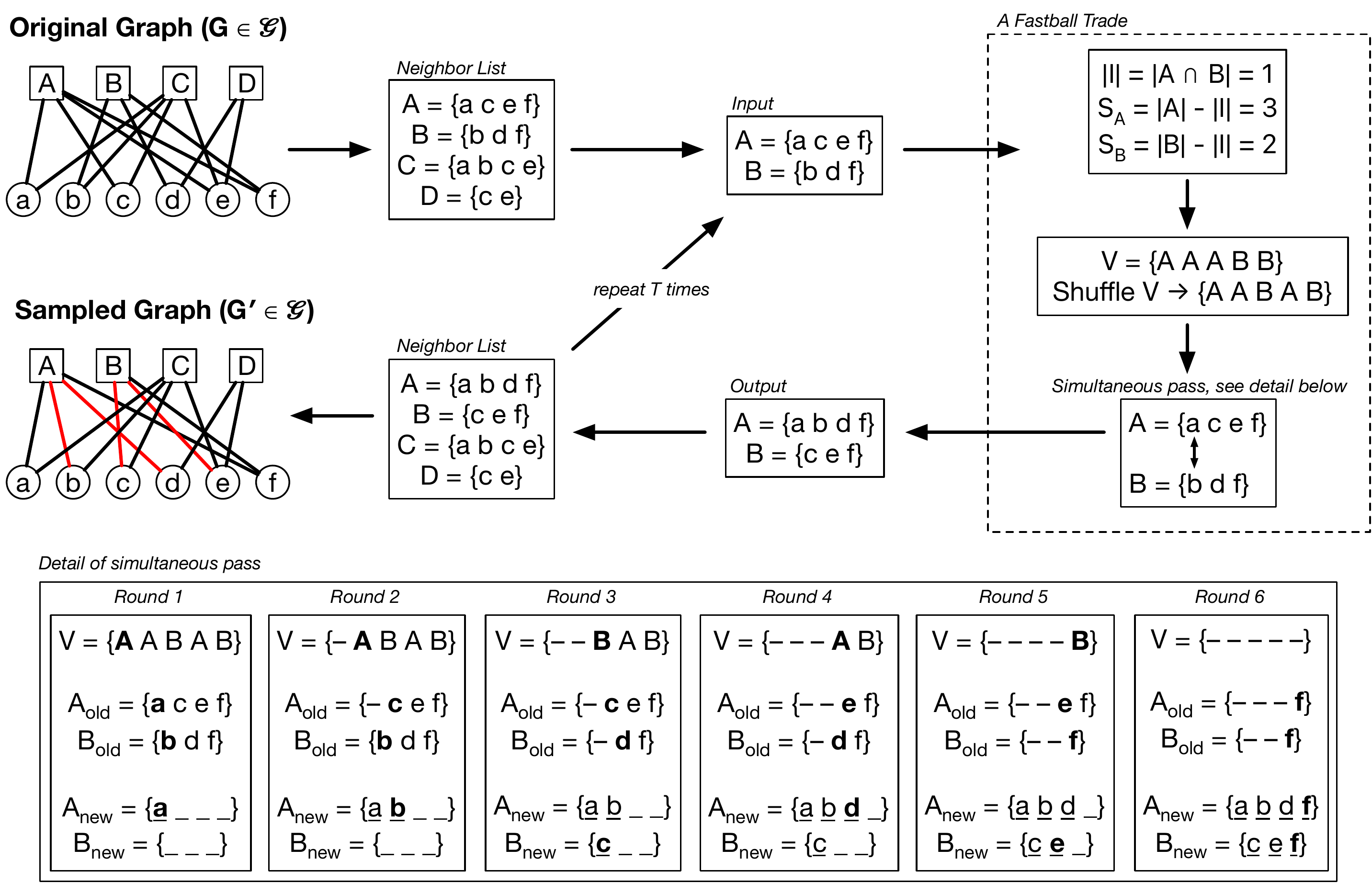}
\end{figure*}

Figure \ref{fig:fastball} outlines the steps of the fastball algorithm, which illustrates how a starting bipartite graph $\mathbf{G}$ is randomized to yield a new graph $\mathbf{G}'$ that is randomly chosen from $\mathcal{G}$. Like curveball, $\mathbf{G}$ is represented as an adjacency list, from which two top nodes are randomly chosen to participate in a trade, and which later rejoin the complete adjacency list after the trade. Here, we focus only on performing the fastball trade itself (the steps enclosed in the dashed line).

First, we compute the cardinality of $i$ and $j$'s intersection (e.g. child A and B have 1 card in common), and from this compute $i$'s and $j$'s number of unique neighbors, $S_i$ and $S_j$ (e.g., child A has 3 unique cards and child B has 2 unique cards). Second, we construct a vector $V$ that contains $S_i$ $i$s and $S_j$ $j$s, and shuffle it. Finally, we make a single simultaneous pass through $i$ and $j$, comparing the elements. When two elements match, they are retained in both lists (e.g., Round 6). When two elements differ, the lower-valued element becomes adjacent to the node identified in $V$ (e.g., Rounds 1-5).

Using a card trading metaphor, this process mirrors the two children holding their alphabetically sorted cards face down in a deck. Each child turns over the top card. If the face up cards match, both children get to keep them. If the face-up cards do not match, then the child identified by the `victory vector' $V$ `wins' the alphabetically-earlier face up card, while the losing child turns a new card face up. For example, in Round 3 of figure \ref{fig:fastball}, child A turns card c face up and child B turns card d face up. In this round, child B wins card c because they are identified by the current element of the victory vector.

\begin{algorithm}[H]
\caption{Fastball Trade algorithm, $O(n)$}
\label{alg:fastball}
\begin{algorithmic}
\STATE \textbf{Input:} Sorted vector of $i$'s neighbors $N_i$ \& Sorted vector of $j$'s neighbors $N_j$\\
\STATE \textbf{Output:} Sorted vector of $i$'s neighbors $N'_i$ \& Sorted vector of $j$'s neighbors $N'_j$\\
\STATE 
\STATE Let $|I| = |N_i \cap N_j|$ \hfill $O(n)$\\
\STATE Let $V$ be a vector of $|N_i| - |I|$ $i$s and $|N_j| - |I|$ $j$s \hfill $O(n)$\\
\STATE Shuffle $V$ \hfill $O(n)$\\
\STATE 
\STATE Let $a = 0$, $b = 0$, $c = 0$ \hfill Explicit steps of~~\\
\WHILE [\hfill a simultaneous pass] {$a \neq |N_i|$ and $b \neq |N_j|$} 
    \IF[\hfill through $N_i$ and $N_j$,]{$N_i[a] = N_j[b]$} 
        \STATE Append $N_i[a]$ to $N'_i$ \hfill collectively $O(n)$~~\\
        \STATE Append $N_j[b]$ $N'_j$\\
        \STATE Increment a and b\\
    \ELSIF{$N_i[a] < N_j[b]$} 
        \STATE Append $N_i[a]$ to $N'_{V[c]}$\\
        \STATE Increment a and c\\
    \ELSIF{$N_i[a] > N_j[b]$} 
        \STATE Append $N_j[b]$ to $N'_{V[c]}$\\
        \STATE Increment b and c\\
    \ENDIF
\ENDWHILE
\STATE \textbf{if} {$a \neq |N_i|$ and $c \neq |V|$} \textbf{then} Append $N_i[a...|N_i|]$ to $N_{V[c...|V|]}$\\
\STATE \textbf{if} {$b \neq |N_j|$ and $c \neq |V|$} \textbf{then} Append $N_j[b...|N_j|]$ to $N_{V[c...|V|]}$\\
\end{algorithmic}
\end{algorithm}

Algorithm \ref{alg:fastball} formalizes the process for performing a fastball trade, and shows the time complexity of each step. The \textit{while loop} and two \textit{if} commands describe the explicit steps of the simultaneous pass through the $N_i$ and $N_j$ adjacency lists, which collectively can be performed in $O(n)$ time when $N_i$ and $N_j$ are sorted. Therefore, like curveball, the efficiency of fastball's key steps requires that the input vectors be sorted, and that the output vectors be sorted in preparation for the next trade. However, unlike curveball's approach, fastball's simultaneous pass approach ensures that the new adjacency lists $N'_i$ and $N'_j$ are assembled in sorted order, and therefore eliminates the computationally costly need to sort them later. Despite this algorithmic modification, the outcome of fastball trades are identical to those performed by curveball, which ensures that fastball also samples $\mathbf{G}~\in~\mathcal{G}$ uniformly at random \cite{carstens2015proof,mitzenmacher2017probability}.

\section{Results}
\label{sec:results}
\subsection{Practical Running Time}
\label{sec:time}
To compare the running times of curveball and fastball, we first implemented each algorithm in C\texttt{++}. By using a low-level language, as opposed to a higher-level language such as R or Python, we are able to more closely match the theoretical time complexities shown in Algorithms \ref{alg:curveball} and \ref{alg:fastball}. We then used these functions to perform 100 trades on bipartite graphs with two top nodes and differing numbers of bottom nodes, where each top node is adjacent to a unique half of the bottom nodes. This type of bipartite graph is likely quite unusual in practice, but is ideal for this experiment for two reasons. First, as Figures \ref{fig:curveball} and \ref{fig:fastball} illustrate, curveball and fastball trades only involve two top nodes; additional top nodes that might be present in the bipartite graph play no role in the (time required for the) trade. Using a bipartite graph with only two top nodes allows us to consider the running time for trades in bipartite graphs with many bottom nodes (here, up to $10^6$), while minimizing the amount of memory needed to hold the graph. Second, ensuring that each top node is adjacent to a unique half of the bottom nodes maximizes the number of possible swaps within a trade, and therefore represents a `worse case scenario' in terms of running time.

\begin{figure}
\caption{Running times of curveball and fastball algorithms}
\label{fig:time}
\centering
\includegraphics[width=.6\columnwidth]{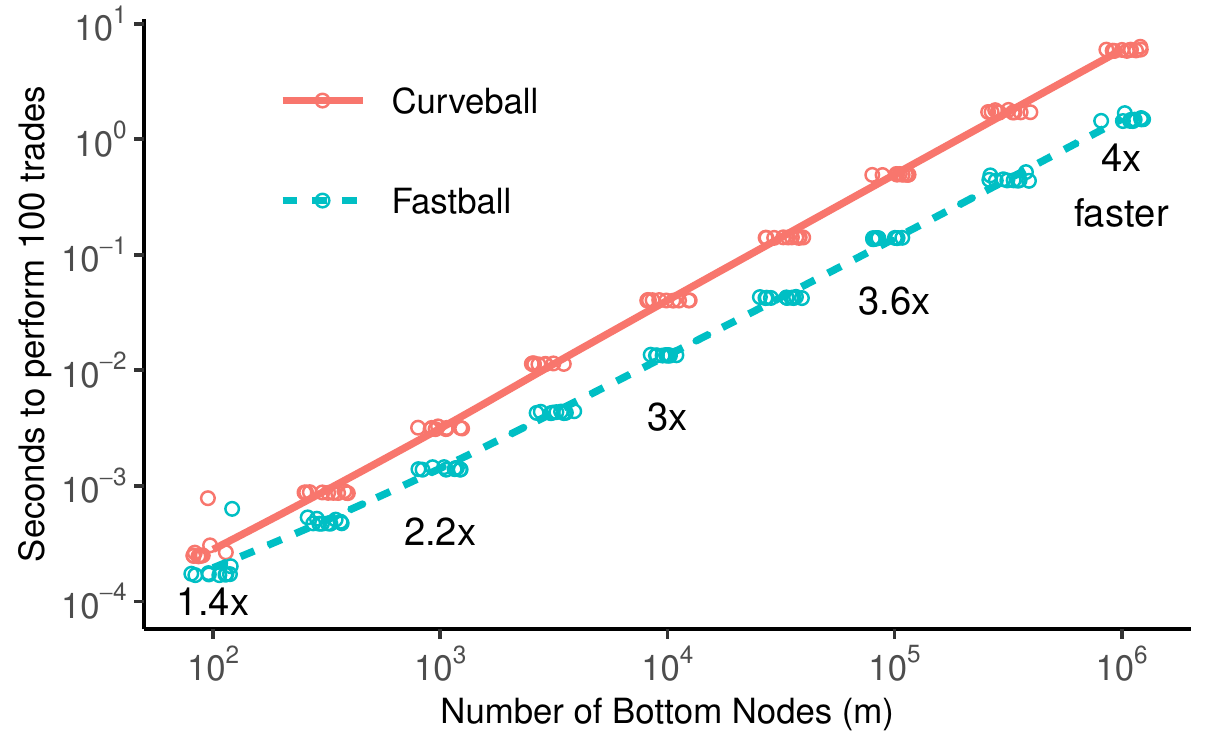}
\end{figure}

Figure \ref{fig:time} shows the time required on an Apple M1 Max processor for curveball (red solid line) and fastball (blue dashed line) to perform 100 trades, as a function of the number of bottom nodes in the bipartite graph. For each number of bottom nodes we performed 10 replications, and for selected numbers we report how much faster fastball is compared to curveball. We find that for these numbers of bottom nodes, fastball is always faster than curveball. As expected given these algorithms' asymptotic time complexities, the improvement in running time offered by fastball is larger when the bipartite graph contains more bottom nodes. For example, fastball is 2.2 times faster in smaller bipartite graphs (e.g., $m = 10^3$) , and four times faster in larger bipartite graphs (e.g. $m = 10^6$).

\subsection{Applying fastball for backbone extraction}
Bipartite graph sampling algorithms have many possible applications. In this section we illustrate one practical application of the fastball algorithm: the fixed degree sequence model for extracting the backbone of bipartite projections. Additionally, this illustration contextualizes the running time improvement offered by fastball when applied to a more realistic bipartite graph than studied in section \ref{sec:time}.

Given a bipartite graph, network researchers often study the bipartite projection, which captures co-occurrences of the bottom nodes among pairs of top nodes. Bipartite projections are weighted, which can complicate their analysis. Additionally, bipartite projections are typically very dense and highly clustered, which can obscure underlying structures \cite{latapy2008basic}. For these reasons, it can be useful to focus on the \textit{backbone} of a bipartite projection, which is an unweighted graph that preserves only the edges whose weights are statistically significant. While many null models exist for determining an edge's statistical significance, the most statistically powerful is the fixed degree sequence model (FDSM), which compares an edge's observed weight in the bipartite projection to the distribution of its weights in the projection of bipartite graphs with the same degree sequences \cite{neal2021comparing,zweig2011systematic}. The shape of this null edge weight distribution is unknown and must be approximated using Monte Carlo methods, which involve repeatedly sampling and constructing the projection of bipartite graphs with given degree sequences. This sampling step is the most computationally costly part of the FDSM, and therefore is the bottleneck for its practical application. By performing the sampling more efficiently, the fastball algorithm improves the practicality of the FDSM as a backbone extraction model \cite{neal2022backbone}.

\begin{figure*}
\caption{Example of bipartite backbone extraction}
\label{fig:backbone}
\centering
\includegraphics[width=.8\textwidth]{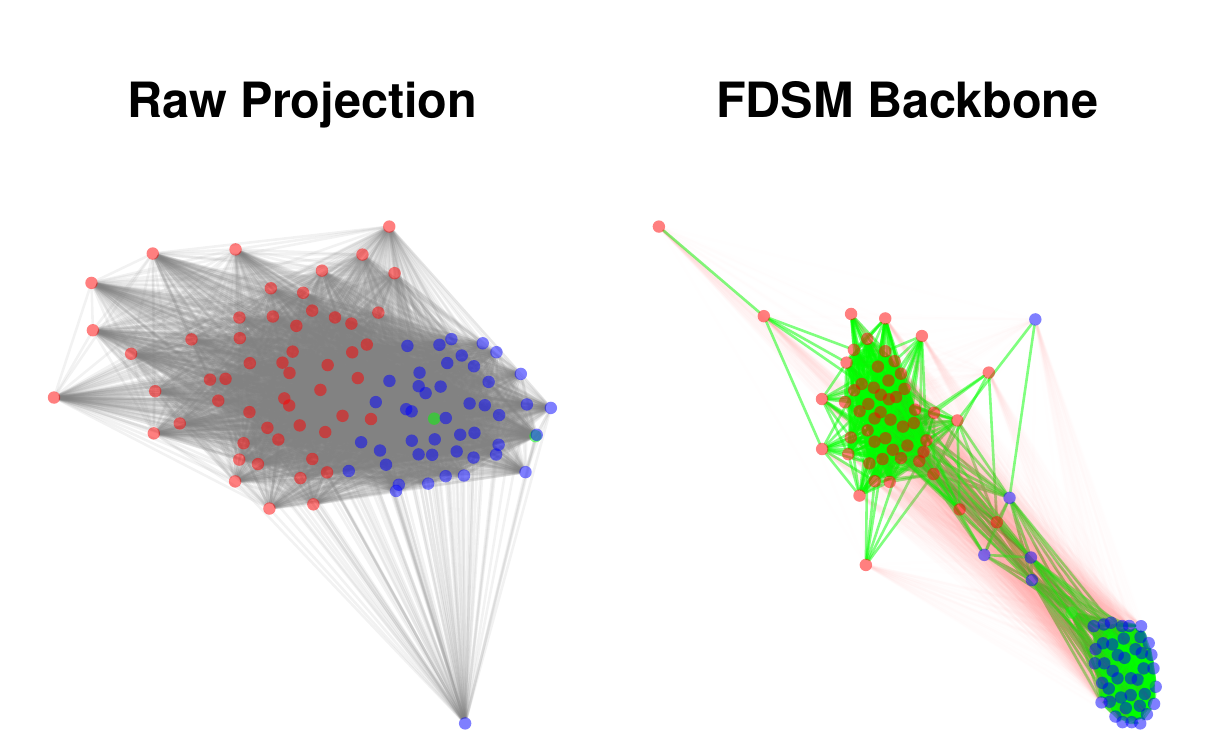}
\end{figure*}

Legislative co-sponsorship networks are one case where bipartite backbone extraction is helpful, and offer an example of the advantages of fastball sampling. In the 116\textsuperscript{th} US Senate, 102 Senators sponsored 5086 bills. These data can be arranged as a bipartite graph where Senators are connected to the bills they sponsored. The top node Senator degree sequence captures the number of bills each Senator sponsored, while the bottom node bill degree sequence captures the number of Senators sponsoring each bill. This bipartite graph can be transformed into a bipartite projection where Senators are connected to other Senators by their number of co-sponsored bills. The left panel of Figure \ref{fig:backbone} illustrates this bipartite projection, with Republican Senators colored red, Democratic Senators colored blue, and Independent Senators colored green. This network is so dense that no underlying structure, including the known polarized structure of the US Senate, can be discerned. Extracting the backbone of this projection can reveal such an underlying structure by preserving only the statistically significant edges. Determining edges' statistical significant under the FDSM requires randomly sampling bipartite graphs with the same degree sequences as the observed $102 \times 5086$ bipartite graph.

Using trade algorithms to perform the sampling required by the FDSM involves making two calculations. First, how many trades are necessary to ensure that the sampled bipartite graph $\mathbf{G'}$ is uniformly drawn from the space of all graphs with the same degrees $\mathcal{G}$? The precise mixing time of trade algorithms is unknown, however numerical experiments have shown that $5n$ is typically sufficient to ensure uniformly random sampling \cite{strona2014fast,carstens2015proof,carstens2018speeding}. Therefore, drawing each sample requires performing $102 \times 5 = 510$ trades. Second, how many samples must be drawn to approximate the null edge weight distribution with sufficient precision to determine whether an observed edge's weight is statistically significantly larger (or smaller) than the null model expectation? The statistical test involved in backbone extraction compares one proportion (an edge's $p$-value under the null model) to another proportion (the chosen $\alpha$ statistical significance level), and therefore the number of required samples is equivalent to the required sample size for comparing two proportions with a given statistical power \cite{neal2021comparing}. In this case, conducting a two-tailed statistical test of an edge's weight at the $\alpha = 0.05$ significance level with 95\% power requires drawing 164,710 samples.

Drawing these samples using our C\texttt{++} implementation of the curveball algorithm takes about 27 minutes on an Apple M1 Max processor. In contrast, drawing these samples using our C\texttt{++} implementation of the fastball algorithm takes only about 11 minutes. These sampling methods yield the same FDSM backbone shown in the right panel of Figure \ref{fig:backbone}, where significantly larger-than-expected edges are drawn in green and significantly smaller-than-expected edges are drawn in red, which clearly highlights the polarized structure of the US Senate. However, the fastball algorithm allows the extraction of an FDSM backbone substantially faster, thereby making the analysis of such data more practical.

\section{Discussion}
\label{sec:discussion}
The fastball algorithm represents a more theoretically efficient and practically faster variant of the existing curveball algorithm. Theoretically, we show that fastball has an asymptotic time complexity of $O(n)$, making it more computationally efficient than curveball, which has an asymptotic time complexity of $O(n~log~n)$. Practically, we show that when implemented in a low-level language such as C\textit{++}, the fastball algorithm performs trades faster than curveball. Specifically, we show that it draws the samples necessary to extract the backbone of a US Senate co-sponsorship network using the FDSM roughly 2.5 times faster, and can perform trades in very large bipartite graphs ($m = 10^6)$ roughly four times faster.

Like curveball, fastball has applications in a range of fields including network science, mathematics, ecology, thermodynamics, and statistical physics \cite{barre2007ensemble,touchette2015equivalence,squartini2015breaking,cimini2019statistical,barvinok2010number,bruno2020ambiguity,gotelli2000null,neal2014backbone}. However, it is important to be clear about their relationship and relative advantages. While both fastball and curveball are trade-type randomization algorithms, fastball might be regarded as an efficient variant of curveball, rather than as an entirely novel algorithm. Curveball might be preferable for the randomization of a small bipartite graph $\mathbf{G}$, or for drawing a small number of samples from $\mathbf{G}$, because it is only negligibly slower than fastball, but can be efficiently implemented in high-level languages such as R and Python that are easier to understand and modify. In contrast, while fastball is also suitable in such cases, it is most useful for randomizing larger bipartite graphs or drawing many samples, but requires implementation in a lower-level language such as C\texttt{++}. However, the R function \texttt{fastball()} in the \texttt{backbone} package \cite{neal2022backbone} provides a user-friendly wrapper that does not require knowledge of C\texttt{++}, and can be used seamlessly with other R functions.

In addition to its practical applications, the fastball algorithm also offers a starting point for both theoretical and algorithmic future directions. First, relatively little is known about trade-type algorithms' mixing times. Because fastball can perform trades faster than curveball, it can be used to extend earlier numerical experiments of mixing time on larger matrices or matrices with unique structures \cite{carstens2015proof,carstens2018speeding}. Second, still greater computational efficiency may be achievable by combining fastball's efficient trade algorithm with the parallelization and I/O-efficient techniques that have been applied to curveball \cite{carstens2018unifying,carstens_et_al}.

\section*{Data Availability}
Algorithm implementations in C\texttt{++}, and R code to reproduce the analyses, are available at \url{https://github.com/zpneal/fastball}.

\section*{Contributor Statement}
KG conceptualized the fastball algorithm, and implemented both algorithms in C\texttt{++}; ZPN revised the algorithms and conducted the analysis. Both authors drafted and revised the paper.

\section*{Acknowledgements}
This work was supported by the National Science Foundation (\#2016320 and \#2211744).


\begin{thebibliography}{00}

\bibitem{admiraal2008networksis}
Admiraal, R. {\&} Handcock, M.~S. (2008)  Networksis: a package to simulate
  bipartite graphs with fixed marginals through sequential importance sampling.
  {\em Journal of {S}tatistical {S}oftware}, \textbf{24}(8).

\bibitem{barre2007ensemble}
Barr{\'e}, J. {\&} Gon{\c{c}}alves, B. (2007)  Ensemble inequivalence in random
  graphs. {\em Physica A: Statistical Mechanics and its Applications},
  \textbf{386}(1), 212--218.

\bibitem{barvinok2010number}
Barvinok, A. (2010)  On the number of matrices and a random matrix with
  prescribed row and column sums and 0--1 entries. {\em Advances in
  {M}athematics}, \textbf{224}(1), 316--339.

\bibitem{bezakova2007sampling}
Bez{\'a}kov{\'a}, I., Bhatnagar, N. {\&} Vigoda, E. (2007)  Sampling binary
  contingency tables with a greedy start. {\em Random {S}tructures \&
  {A}lgorithms}, \textbf{30}(1-2), 168--205.

\bibitem{blanchet2013characterizing}
Blanchet, J. {\&} Stauffer, A. (2013)  Characterizing optimal sampling of
  binary contingency tables via the configuration model. {\em Random
  {S}tructures \& {A}lgorithms}, \textbf{42}(2), 159--184.

\bibitem{boroojeni2017generating}
Boroojeni, A.~A., Dewar, J., Wu, T. {\&} Hyman, J.~M. (2017)  Generating
  bipartite networks with a prescribed joint degree distribution. {\em Journal
  of {C}omplex {N}etworks}, \textbf{5}(6), 839--857.

\bibitem{bruno2020ambiguity}
Bruno, M., Saracco, F., Garlaschelli, D., Tessone, C.~J. {\&} Caldarelli, G.
  (2020)  The ambiguity of nestedness under soft and hard constraints. {\em
  Scientific {R}eports}, \textbf{10}(1), 1--13.

\bibitem{carstens2015proof}
Carstens, C.~J. (2015)  Proof of uniform sampling of binary matrices with fixed
  row sums and column sums for the fast curveball algorithm. {\em Physical
  {R}eview {E}}, \textbf{91}(4), 042812.

\bibitem{carstens2018unifying}
Carstens, C.~J., Berger, A. {\&} Strona, G. (2018a)  A unifying framework for
  fast randomization of ecological networks with fixed (node) degrees. {\em
  MethodsX}, \textbf{5}, 773--780.

\bibitem{carstens_et_al}
Carstens, C.~J., Hamann, M., Meyer, U., Penschuck, M., Tran, H. {\&} Wagner, D.
  (2018b)  {Parallel and I/O-efficient Randomisation of Massive Networks using
  Global Curveball Trades}. In Azar, Y., Bast, H. {\&} Herman, G., editors,
  {\em 26th Annual European Symposium on Algorithms (ESA 2018)}, volume 112 of
  {\em Leibniz International Proceedings in Informatics (LIPIcs)}, pages
  11:1--11:15, Dagstuhl, Germany. Schloss Dagstuhl--Leibniz-Zentrum fuer
  Informatik.

\bibitem{carstens2018speeding}
Carstens, C.~J. {\&} Kleer, P. (2018)  Speeding up switch Markov chains for
  sampling bipartite graphs with given degree sequence. In {\em Approximation,
  Randomization, and Combinatorial Optimization. Algorithms and Techniques
  (APPROX/RANDOM 2018)}. Schloss Dagstuhl-Leibniz-Zentrum fuer Informatik.

\bibitem{chen2006sequential}
Chen, Y., Dinwoodie, I.~H. {\&} Sullivant, S. (2006)  Sequential importance
  sampling for multiway tables. {\em The {A}nnals of {S}tatistics},
  \textbf{34}(1), 523--545.

\bibitem{cimini2019statistical}
Cimini, G., Squartini, T., Saracco, F., Garlaschelli, D., Gabrielli, A. {\&}
  Caldarelli, G. (2019)  The statistical physics of real-world networks. {\em
  Nature Reviews Physics}, \textbf{1}(1), 58--71.

\bibitem{gale1957theorem}
Gale, D.  et~al. (1957)  A theorem on flows in networks. {\em Pacific {J}ournal
  of {M}athematics}, \textbf{7}(2), 1073--1082.

\bibitem{gotelli2000null}
Gotelli, N.~J. (2000)  Null model analysis of species co-occurrence patterns.
  {\em Ecology}, \textbf{81}(9), 2606--2621.

\bibitem{latapy2008basic}
Latapy, M., Magnien, C. {\&} Del~Vecchio, N. (2008)  Basic notions for the
  analysis of large two-mode networks. {\em Social networks}, \textbf{30}(1),
  31--48.

\bibitem{mitzenmacher2017probability}
Mitzenmacher, M. {\&} Upfal, E. (2017) {\em Probability and computing:
  Randomization and probabilistic techniques in algorithms and data analysis}.
Cambridge university press.

\bibitem{neal2014backbone}
Neal, Z.~P. (2014)  The backbone of bipartite projections: Inferring
  relationships from co-authorship, co-sponsorship, co-attendance and other
  co-behaviors. {\em Social {N}etworks}, \textbf{39}, 84--97.

\bibitem{neal2022backbone}
Neal, Z.~P. (2022)  backbone: An R package to extract network backbones. {\em
  PloS one}, \textbf{17}(5), e0269137.

\bibitem{neal2021comparing}
Neal, Z.~P., Domagalski, R. {\&} Sagan, B. (2021)  Comparing Alternatives to
  the Fixed Degree Sequence Model for Extracting the Backbone of Bipartite
  Projections. {\em Scientific {R}eports}.

\bibitem{penschuck2020recent}
Penschuck, M., Brandes, U., Hamann, M., Lamm, S., Meyer, U., Safro, I.,
  Sanders, P. {\&} Schulz, C. (2020)  Recent advances in scalable network
  generation. {\em arXiv preprint arXiv:2003.00736}.

\bibitem{ryser1957combinatorial}
Ryser, H.~J. (1957)  Combinatorial properties of matrices of zeros and ones.
  {\em Canadian {J}ournal of {M}athematics}, \textbf{9}, 371--377.

\bibitem{squartini2015breaking}
Squartini, T., de~Mol, J., den Hollander, F. {\&} Garlaschelli, D. (2015)
  Breaking of ensemble equivalence in networks. {\em Physical Review Letters},
  \textbf{115}(26), 268701.

\bibitem{strona2014fast}
Strona, G., Nappo, D., Boccacci, F., Fattorini, S. {\&} San-Miguel-Ayanz, J.
  (2014)  A fast and unbiased procedure to randomize ecological binary matrices
  with fixed row and column totals. {\em Nature {C}ommunications},
  \textbf{5}(1), 1--9.

\bibitem{touchette2015equivalence}
Touchette, H. (2015)  Equivalence and nonequivalence of ensembles:
  thermodynamic, macrostate, and measure levels. {\em Journal of Statistical
  Physics}, \textbf{159}(5), 987--1016.

\bibitem{verhelst2008efficient}
Verhelst, N.~D. (2008)  An efficient MCMC algorithm to sample binary matrices
  with fixed marginals. {\em Psychometrika}, \textbf{73}(4), 705--728.

\bibitem{zweig2011systematic}
Zweig, K.~A. {\&} Kaufmann, M. (2011)  A systematic approach to the one-mode
  projection of bipartite graphs. {\em Social {N}etwork {A}nalysis and
  {M}ining}, \textbf{1}(3), 187--218.

\end{thebibliography}
\end{document}